\documentclass[aps,prb,twocolumn,superscriptaddress,showpacs]{revtex4}

\usepackage[utf8]{inputenc}
\usepackage[english]{babel}
\usepackage{graphicx,tipa}
\usepackage{amsmath}
\usepackage{amssymb}
\usepackage{xcolor}
\usepackage{color,wasysym}
\usepackage{bbold}
\usepackage[abs]{overpic}

\usepackage{cancel}

\definecolor{redred}{HTML}{D53E4F}

\definecolor{greengreen}{HTML}{2F9B31}

\newcommand{\Id}{{\mathbb 1}}
\newcommand{\trace}{\mbox{Tr}}
\newcommand{\ie}{i.e.~}

\newcommand{\temp}{T}%{\kappa T}	% Temperature. Eventually remove Boltzmann constant from here (Psi likes it though).
\newcommand{\free}{{\cal F}}		% Free energy. Calligraphic F as Saro likes.
\newcommand{\DOS}{\text{D}}	% The DOS. Exponentially extensive.
\newcommand{\ense}{\rho}

\newcommand{\bDOS}{D_{\eta}}%{\text{b-}\DOS}

\newcommand{\HSTNTitleText}{Density-of-states of many-body quantum systems from tensor networks}
\begin{document}

\title{\HSTNTitleText}	% change in HSTN_preamble.tex

\author{Fabian Schrodi}
\affiliation{Institute for Complex Quantum Systems \& Center for Integrated Quantum Science and Technologies, Universit\"at Ulm, D-89069 Ulm, Germany}

\author{Pietro Silvi}
\affiliation{Institute for Complex Quantum Systems \& Center for Integrated Quantum Science and Technologies, Universit\"at Ulm, D-89069 Ulm, Germany}
\affiliation{Institute for Theoretical Physics, University of Innsbruck, A-6020 Innsbruck, Austria}

\author{Ferdinand Tschirsich}
\affiliation{Institute for Complex Quantum Systems \& Center for Integrated Quantum Science and Technologies, Universit\"at Ulm, D-89069 Ulm, Germany}

\author{Rosario Fazio}
\affiliation{ICTP, Strada Costiera 11, 34151 Trieste, Italy}
\affiliation{NEST, Scuola Normale Superiore and Istituto Nanoscienze-CNR, I-56126, PISA, Italy}

\author{Simone Montangero}
\affiliation{Theoretische Physik, Universit\"at des Saarlandes, D-66123 Saarbr\"ucken, Germany}
\affiliation{Institute for Complex Quantum Systems \& Center for Integrated Quantum Science and Technologies, Universit\"at Ulm, D-89069 Ulm, Germany}

%\author{Other Contributor}
%\affiliation{Affiliation of the other contributor}
\date{\today}

\begin{abstract}
We present a technique to compute the microcanonical thermodynamical properties of a many-body quantum system using tensor networks.
The Density Of States (DOS), and more general spectral properties,
are evaluated by means of  a
Hubbard-Stratonovich transformation performed on top of a real-time evolution, which is carried out via numerical methods based on tensor networks.
As a consequence, the free energy and thermal averages can be also calculated.
We test this approach on the one-dimensional Ising and Fermi-Hubbard models.
Using matrix product states, we show that  
the thermodynamical quantities as a function of temperature are in very good agreement with the exact results.
%This approach, in principle, can be extended to higher-dimensional systems by properly employing other types of tensor networks. 
\end{abstract}

\maketitle

Understanding the properties of strongly correlated quantum many-body systems is one of the main challenges of modern physics, 
with implications ranging from  condensed matter to quantum technologies and high-energy physics. To this aim several different 
powerful numerical techniques such  as Monte Carlo or different variants of numerical renormalisation and variational approaches 
(see e.g.~\cite{Anderson-book,Bulla2008, Schollwock2005}) have been developed in order to explore regimes unaccessible to 
analytical approaches. Each of those methods has advantages and shortcomings, all together they offer a  solid platform for the 
investigation of strongly correlated phenomena.  Despite the tremendous advancements experienced in the last decades by both 
theoretical and numerical approaches~\cite{momentsDOS}, the complete access to the spectral
properties of a many-body quantum system 
is a long-sighted goal which remained elusive in those cases where accurate Monte Carlo
sampling is not possible (due, for example, to sign problems~\cite{free-MC1,free-MC2,rago}).
Having access to the many-body density of states (DOS) and energy-resolved observable quantities, thus effectively reconstructing its 
microcanonical ensemble, grants in turn full knowledge of its free-energy, and thus of the whole thermodynamics. 

Here we introduce a technique, based on tensor network ansatz states, to compute the microcanonical density of states of a many-body 
quantum system, and from there to have a direct access to the free-energy of the system. During the last decades Tensor Networks 
(TNs) methods have proven to be one of the most promising approaches to low-dimensional quantum systems, both in and out of 
equilibrium~\cite{SchollwockAGEofMPS,orusrev}. The method we describe exploits two key features of TN techniques: the ability to encode efficiently 
a many-body quantum state and to simulate its (real-time) dynamics. We will show that a smoothened version of the microcanonical 
density of states can be evaluated using the Hubbard-Stratonovich transformation~\cite{Strat0,Hubb0} and a subsequent time evolution 
of the system. We benchmark this approach against two paradigmatic models: the Ising model in transverse field and the Hubbard model. 
Here we focus on one-dimensional systems, however the method is in principle not bound to this limitation, provided an appropriate TN 
ansatz state is used.

TN methods are based on an efficient encoding of the many-body quantum state that allows simulations of large system sizes, unattainable via 
direct diagonalisation, while at the same time guaranteeing a remarkable precision. A vast class of TN algorithms  have  focused on an accurate 
determination of ground-state properties~\cite{SchollwockAGEofMPS,orusrev}. Prominent examples beyond the density matrix renormalisation 
group~\cite{WhiteDMRG}, the workhorse and inspiration of TN methods,  are algorithms based on different one- and two-dimensional tensor network 
structures like Matrix Product States (MPS)~\cite{MPSZero,MPSOne,SchollwockAGEofMPS}, MERA~\cite{VidalMERA}, PEPS~\cite{PEPSa}  or tree-TN~\cite{MatthiasTTN}. Their main common feature is the ability to minimise 
an energy functional over a tailored variational class of states and hence reconstruct, for instance, the many-body ground states of lattice Hamiltonians~\cite{MPSOne,PEPSb},
and some elementary excitations \cite{MPSexcite,MPSexciteB}.
TN techniques have been successfully extended to describe the evolution of isolated~\cite{VidalTEBD,daley04,TDMRG} and open quantum systems~\cite{Zwolak,MPDO,LPTN}.
The same techniques can be repurposed to reconstruct many-body states
at finite temperature \cite{MPDO, METTS,ThermoPEPS1,ThermoSchwinger,SETTN}, thus proving that tensor networks are tools also capable of reconstructing efficiently canonical (Gibbs) ensembles \cite{METTS2}.

We introduce the method in Sec.~\ref{sec:method}.
Also, we put forward two proposals on how to
adapt the standard tensor network methods for real-time evolution \cite{VidalTEBD, TDMRG} to this purpose.
In Sec.~\ref{sec:results}, we test the approach on two paradigmatic models in one-dimension.
Additionally, we show how the method can be extended to compute the microcanonical average of a generic 
observable. Finally, we draw the concluding remarks in Sec.~\ref{sec:conc}.

%\textit{The method} $-$
\section{Method} \label{sec:method}
The calculation of the Density of States $\DOS(E) = \sum_{\nu} \delta(E - E_{\nu})$ requires 
in principle   the whole spectrum of the Hamiltonian $H$, as the sum $\sum_\nu$ runs over all the many-body eigenstates $|\nu\rangle$ of $H$ 
(with $H |\nu\rangle = |\nu\rangle E_\nu$). Here we show how to reconstruct, via TNs simulations, the $\DOS(E)$ without 
the need to explicitly diagonalise $H$, {in the same spirit of Ref.~\cite{Osborne}}. As a side note, once $\DOS(E)$ is known,
the free energy $\free$
can be easily extracted via $\free(T) = - \temp \ln \int e^{-E/\temp} \DOS(E) dE$, where
$T$ is the temperature (we set $k_B=\hbar=1$).

We consider a Gaussian broadening of the Dirac delta $\delta(x) \to  \delta_{\eta}(x) := \sqrt{\frac{\eta}{\pi}} e^{-\eta x^2}$, 
which satisfies $\lim_{\eta \to \infty} \delta_{\eta}(x) = \delta(x)$ in the sense of distributions. Accordingly, we define a broadened density of states
$\bDOS(E) = \sqrt{\frac{\eta}{\pi}}  \sum_\nu e^{-\eta (E - E_\nu)^2} = \sqrt{\frac{\eta}{\pi}} \trace[ e^{-\eta (H - E)^2} ]$. 
The corresponding `canonical' quantity $\free_{\eta}(T) = - \temp \ln \int e^{-E/\temp} \bDOS(E) dE$ is a good approximation of the free energy of the 
system $\free_{\eta} (T) \approx \free(T)$ as long as the  broadening is sharper than the exponential decay introduced by temperature, 
\ie $\temp \gg \eta^{-1/2}$. An important check in this respect is to guarantee that the $\bDOS(E)$ preserves the spectral measure, i.e. that 
$\int \bDOS(E) dE = \int \DOS(E) dE$, which is, in turn, equal to the total number of states ($d^L$ where $d$ is the local space dimension and $L$ is the system size).

By means of Fourier analysis, it is possible to rewrite
\begin{equation}
 \sqrt{\frac{\eta}{\pi}} e^{-\eta(H - E)^2} = \frac{1}{2\pi} \int_{-\infty}^{+\infty} e^{-\frac{t^2}{4\eta}+itE} e^{-itH} \, dt,
\end{equation}
usually known as Hubbard-Stratonovich transformation~\cite{Strat0,Hubb0}.
This effectively recasts the problem of characterising the DOS into a problem of simulating the real-time evolution of the 
quantum system. Indeed, the $\bDOS$ can be obtained via the numerical integral
\begin{equation} \label{eq:bdosintegral}
\bDOS(E) = \frac{1}{\pi} \int_{0}^{+\infty} e^{-\frac{t^2}{4\eta}} \,\mathfrak{Re}\!\left( e^{itE} \,\trace\!\left[ e^{-itH}\right] \right) dt,
\end{equation}
which represents the core of the Hubbard-Stratonovich Tensor Network (HS-TN) method introduced here. 
Notice that the Gaussian modulation $e^{-\frac{t^2}{4\eta}}$ drastically reduces the relevance of the long-time contributions
to the $\bDOS$.
As a consequence, it is sufficient to compute $\trace\!\left[ e^{-itH}\right]$ for reasonably short times $t$, typically within a few orders of magnitude of $\sqrt{\eta}$ (fine-tuning details of the time-integration are discussed in Appendix~\ref{app:finetune}.
This enables in practice the HS approach with tensor networks, since short-time evolutions are generally efficient and accurate in TN simulations. 

In this paper we consider one-dimensional quantum systems with open boundary conditions and compute the time evolution using MPS~\cite{MPSZero,MPSOne,SchollwockAGEofMPS} and the Time-Evolving Block Decimation (TEBD) algorithm~\cite{VidalTEBD,TDMRG}.
Alternative techniques to perform real-time evolution of TN, such as the Time-Dependent Variational Principle are similarly viable for this task~\cite{TDVP, TDVPB}.
In order to compute the trace $\trace\!\left[ e^{-itH}\right]$ we propose two alternatives, or pathways:

\begin{figure}
 \begin{center}
 \begin{overpic}[width = \columnwidth, trim={0pt 10pt 0pt 0pt},clip, unit=1pt]{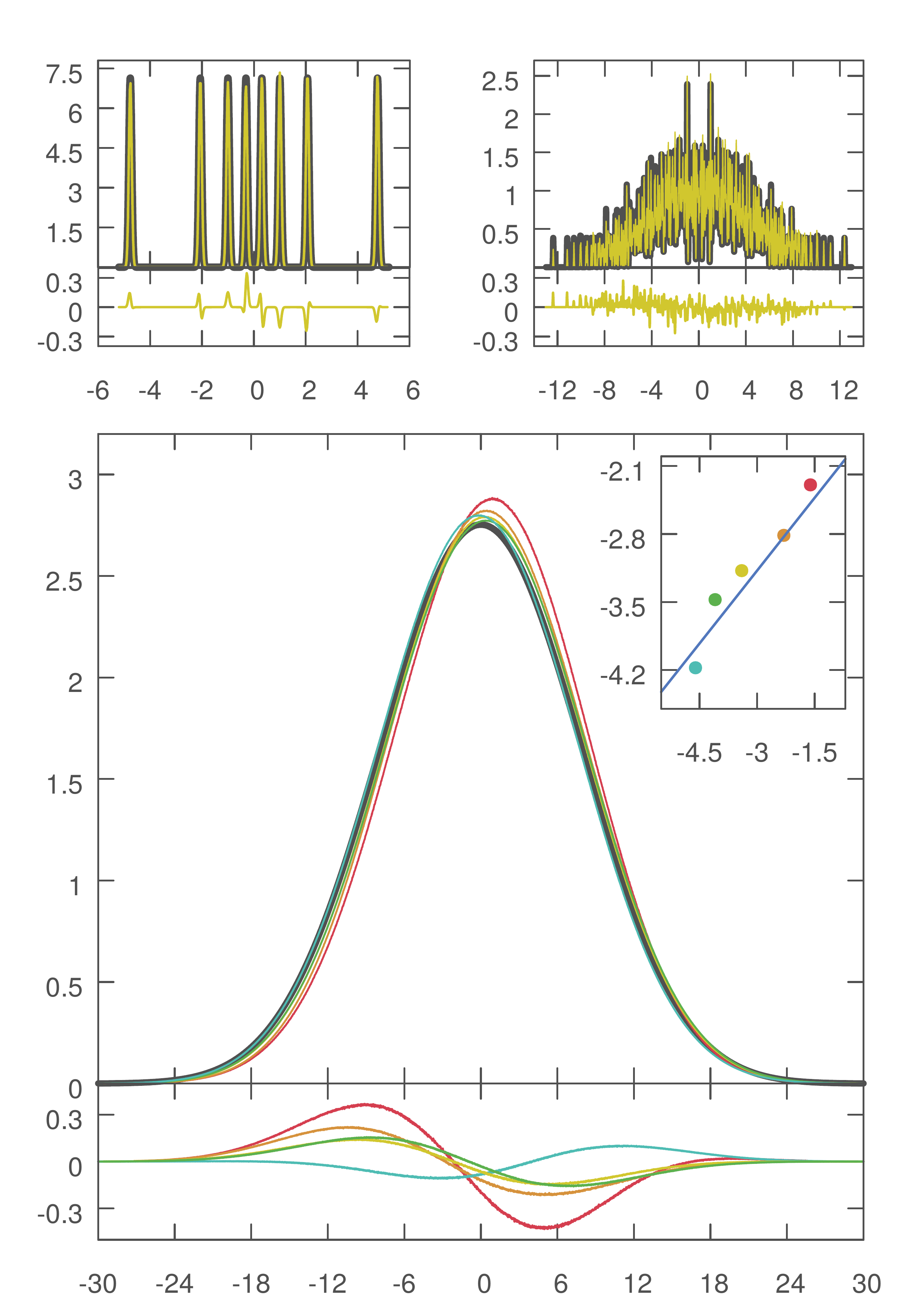}
 \put(0, 210){$D_{\eta,+}$}
 \put(24, 237){\scriptsize $\times 10^{7}$}
 \put(6, 34){$\Delta$}
% \put(-2, 33){$\displaystyle \frac{\Delta}{10^5}$}
 \put(27, 52){\scriptsize $\times 10^{5}$}
 \put(199, -8){$E$}
 \put(-2, 320){$D_{\eta,+}$}
 \put(3, 266){$\Delta$}
 \put(101, 240){$E$} 
 \put(115, 320){$D_{\eta,+}$}
 \put(120, 266){$\Delta$}
 \put(218, 240){$E$} 
 \put(148, 214){$\log \epsilon_0$}
 \put(195, 135){$\log (1/m)$}
  \put(89, 339){\footnotesize $L=4$}
  \put(208, 339){\footnotesize $L=10$}
  \put(40, 221){\footnotesize $L=30$}
 \end{overpic}
\end{center}
\caption{ \label{fig:isingdos} (color online) 
Comparison between the $D_{\eta,+}(E)$ calculated from the exact spectrum (grey curve)
and the $D_{\eta,+}(E)$ calculated via HS-TN method (colored curves). Here we considered the
spin-$\frac{1}{2}$ Ising model in 1+1D, in the even parity symmetry sector
($h = 1, \eta = 10 L^2$) for $L = 4$, $m=16$, $N_R = 1000$ (top left), $L=10$, $m=32$, $N_R = 50$ (top right), while
 $L=30$, $N_R = 20$ and $m = 5$, (red), 10 (orange), 30 (yellow), 60 (green), 100 (cyan) in the bottom panel.
 The lower stripe on each panel shows the deviation curve $\Delta = D_{\eta,+,\text{exact}} - D_{\eta,+,m}$.
Inset: The integrated deviation $\epsilon_0 = d^{1-L} \int | \Delta(E) | dE$ between the exact broadened DOS $D_{\eta,+,\text{exact}}$ and the DOS $D_{\eta,+,m}$ obtained via the HS-TN method as a function of $1/m$ (color code as in the main plot).
The blue line is a power-law fit, corresponding to a power law $\epsilon_0 \propto  m^{-1/2}$.
}
\end{figure}
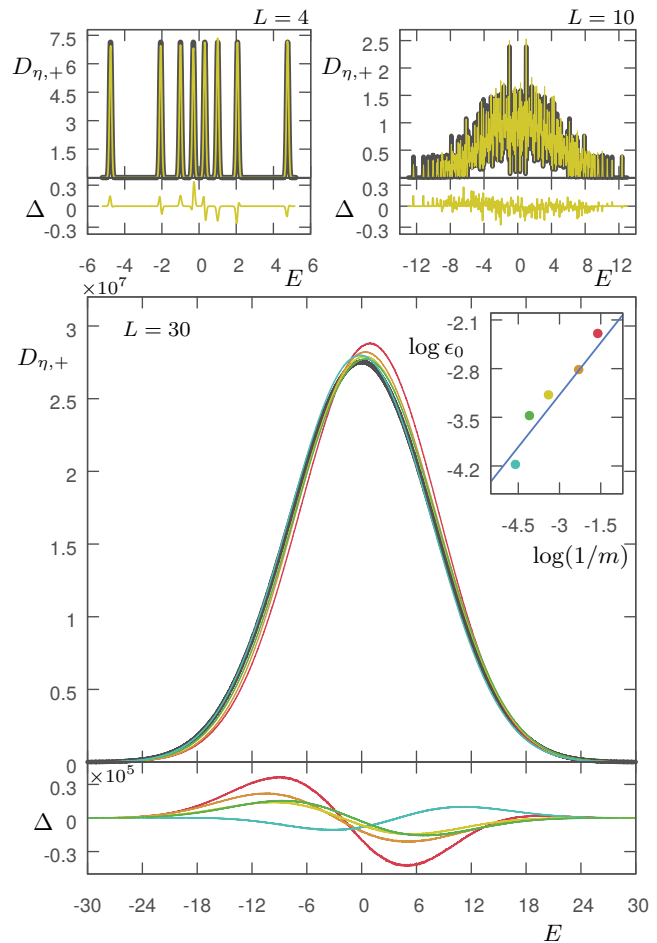

{\it (1)} Random MPS sampling $-$ In this approach, the identity operator in $\trace\!\left[ e^{-itH} \Id \right]$ is 
approximated as the mixture of a large number $N_R$ of random states (see Appendix~\ref{app:repeat} for how to generate them).
$\mathbb{1} \simeq \frac{d^{L}}{N_R} \sum_{R}^{N_R} | \psi_R \rangle \langle \psi_R |$
where  the $| \psi_R \rangle$ are $N_R$ random states. This results in
$\trace\!\left[  e^{itH}\right] = \frac{d^{L}}{N_R} \sum_{R}^{N_R}\langle \psi_R (0) | \psi_R (t) \rangle$,
where  $| \psi_R (t) \rangle = e^{-itH} | \psi_R (0) \rangle = e^{-itH} | \psi_R \rangle$.
The number of random states $N_R$ used to compute the average is increased until the convergence of $\bDOS(E)$ 
is observed. 
This method could be in principle partially biased as the random sampling is limited only to states which contain an 
amount of entanglement upper-bound by $\log m$, where $m$ is the MPS auxiliary bond dimension.  
However, this is not a fundamental limitation since it is conceptually possible to build many-body bases 
composed by states with low entanglement content only (e.g. canonical product bases).

{\it (2)} Local space doubling $-$ We briefly put forward an alternative route to the HS-TN calculation.
In order to calculate the trace of a many-body operator $\trace[A]$,
one can 
double the number of sites and consider the relation $\langle \Phi^+|^{\otimes L} (A_{\text{odd}} \otimes \Id_{\text{even}}) | 
\Phi^{+} \rangle^{\otimes L} = \trace[A]$, where $A$ acts only the odd sites, and
$| \Phi^{+} \rangle^{\otimes L} = \bigotimes_{j = 1}^{L} (\sum_{s = 1}^{d} |s\rangle_{2j-1} |s\rangle_{2j} )$.
The (non-normalised) state $| \Phi^{+} \rangle^{\otimes L}$ can be easily expressed as an MPS with bond dimension $1$ and physical dimension $d^2$
(after merging together sites $2j-1$ and $2j$).
This approach does not require a statistical average, however the drawback 
sits in the increase of the local space dimension to $d^2$.
We realize that this route is more similar in spirit
to Refs.~\cite{Osborne,SETTN}, but with the advantage that it uses the same pure-state time-evolution methods as pathway {\it (1)}.

%\textit{Results} $-$
\section{Results} \label{sec:results}

Hereafter we benchmark the presented method applying it to the spin-$\frac{1}{2}$ Ising model in a transverse field
and the spin-$\frac{1}{2}$ Hubbard model, following pathway {\it (1)}.
We address Abelian global symmetries explicitly~\cite{SinghMain,Singh2007}, and this allows us to reconstruct 
sector-specific broadened DOS, where now
$D_{\eta,q}(E) = \sqrt{\frac{\eta}{\pi}}  \sum_{\nu(q)} e^{-\eta (E - E_{\nu(q)})^2}$
for an arbitrary quantum number $q$, and the sum $\sum_{\nu(q)}$ runs over all the eigenstates $|\nu(q)\rangle$ of $H$ with a fixed quantum 
number $q$. Embedding such symmetry content is equivalent to substituting $\trace[ e^{-itH}] \to \trace[e^{-itH} P_q]$ in Eq.~\eqref{eq:bdosintegral}, where
$P_q = \sum_{\nu(q)} |\nu(q)\rangle \langle \nu(q)|$ is the projector over all the $|\nu(q)\rangle$ states.
Notice that $[e^{-itH},P_q] = 0$ since $q$ are good quantum numbers, connected to a symmetry of $H$.
Calculating such trace is relatively straightforward in pathway {\it(1)},
where one needs simply to sample random states $|\psi^{q}_R\rangle$ having quantum number $q$. 
The sector-specific densities of states can be eventually summed up to recover the full DOS, \ie $\bDOS (E) = \sum_q D_{\eta,q}(E)$.

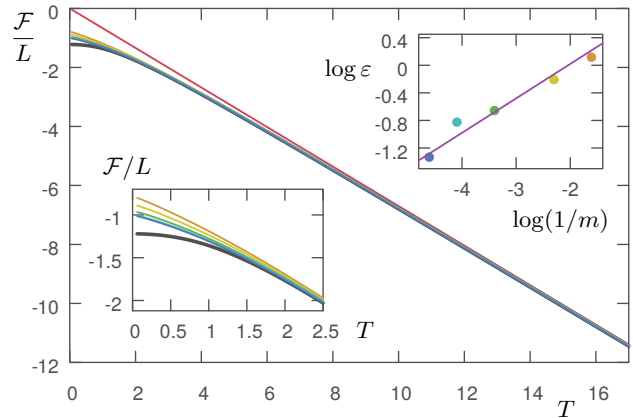
\begin{figure}
 \begin{center}
 \begin{overpic}[width = \columnwidth, unit=1pt]{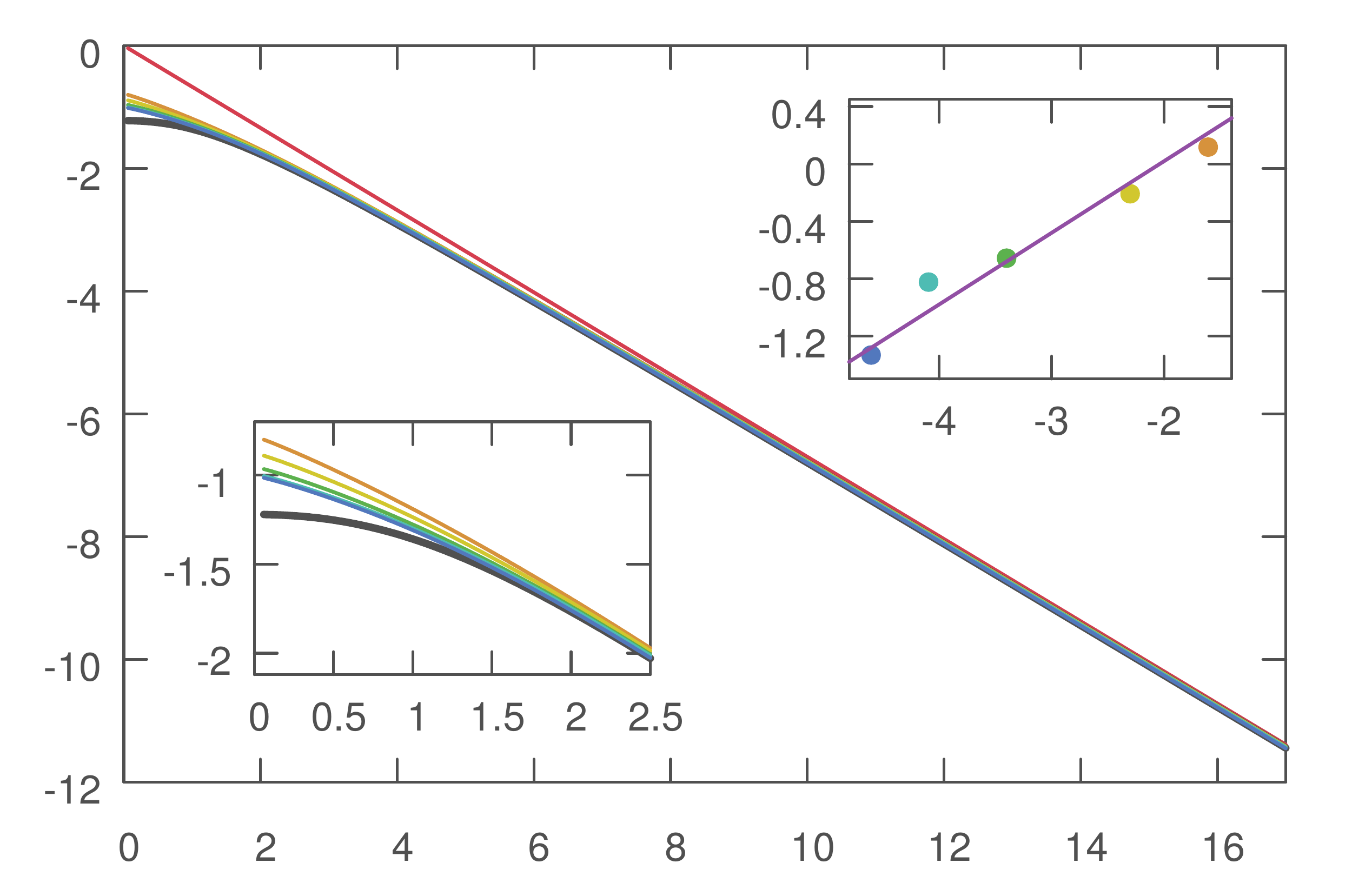}
 \put(0, 141){$\displaystyle\frac{\free}{L}$}
 \put(207, 0){$T$}
 \put(35, 92){$\free/L$}
 \put(131, 30){$T$}
 \put(119, 129){$\log \varepsilon$}
 \put(190, 72){$\log (1/m)$}
 \end{overpic}
%
 %\includegraphics[width = \columnwidth]{Figure1.jpg}
 %
 %Sample figure
\end{center}
\caption{ \label{fig:thermoising} (color online)
Thermodynamics of the spin-$\frac{1}{2}$ Ising model in 1+1D, for a chain of $L = 30$ spins. 
Main panel: the free energy per site $\free/L$ calculated from the exact solution (black line) and from the HS-TN method for $m$ = 5 (orange), 10 (yellow), 30 (green), 60 (cyan), 100 (blue).
The expected high temperature result $\free(T) \approx - \temp (L-1) \ln 2$ is shown via the red line.
Bottom inset: magnification of the low temperature regime. Top inset: error $\varepsilon(m) := L^{-1} \int |\free_{\text{exact}} - \free_{m}| dT$ as a function of $1/m$.
The purple line is a power-law fit, corresponding to $\varepsilon \propto  m^{-1/2}$.
}
\end{figure}

The spectrum of the  Ising model, $H = -\sum_j^{L-1} \sigma^{x}_j \sigma^{x}_{j+1} + h \sum_{j}^{L} \sigma^{z}_j$,
can be computed exactly since it is a free fermion theory~\cite{isingsolution} ($\sigma^{\alpha}$ are the Pauli matrices, $j$ labels the 
lattice sites, $h$ is the transverse field, and $L$ the number of sites). As a figure of merit for our technique,
we consider the $D_{\eta,+}(E)$, restricted to the even `$+$' sector of the parity symmetry $\Pi = \sigma^{z \otimes L}$.
We compare the $D_{\eta,+}$ calculated from the exact spectrum with the one obtained via the HS-TN method.
Such comparison is shown in Fig.~\ref{fig:isingdos} (using $h =1, \eta = 10 L^2$)
for different lattice sizes. 
Our analysis shows clearly that the HS-TN data converge to the exact solution.
In Fig.~\ref{fig:thermoising} (main panel) we further compare the free energy per site
$\free_{\eta}(T)/L$ calculated with the HS-TN method to
the exact solution. We also show that we recover correctly the large $T$ limit behaviour, which predicts $\free(T) \approx - \temp (L-1) \ln 2 + 2^{1-L}\trace[ H ]$ (where $2^{L-1}$ is the dimension of the whole even symmetry sector).
We observe numerical convergence with increasing bond dimension $m$, which
we report in the top inset of Fig.~\ref{fig:thermoising} by plotting the deviation in $\mathcal{L}_1$ norm.
Convergence with respect to the ensemble average $N_R$ is reported in Appendix~\ref{app:repeat}.

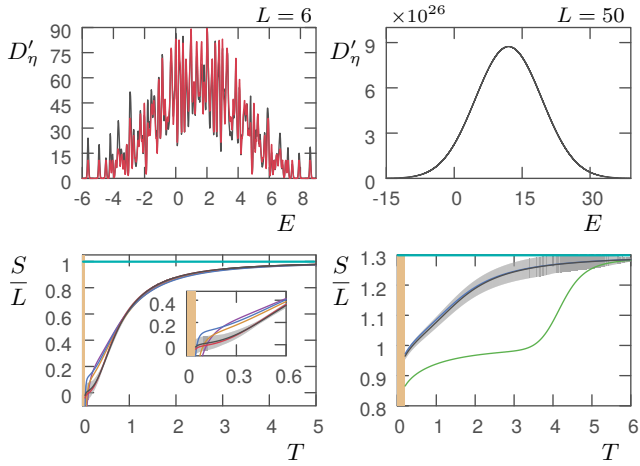
\begin{figure}
 \begin{center}
 \begin{overpic}[width = \columnwidth,  unit=1pt]{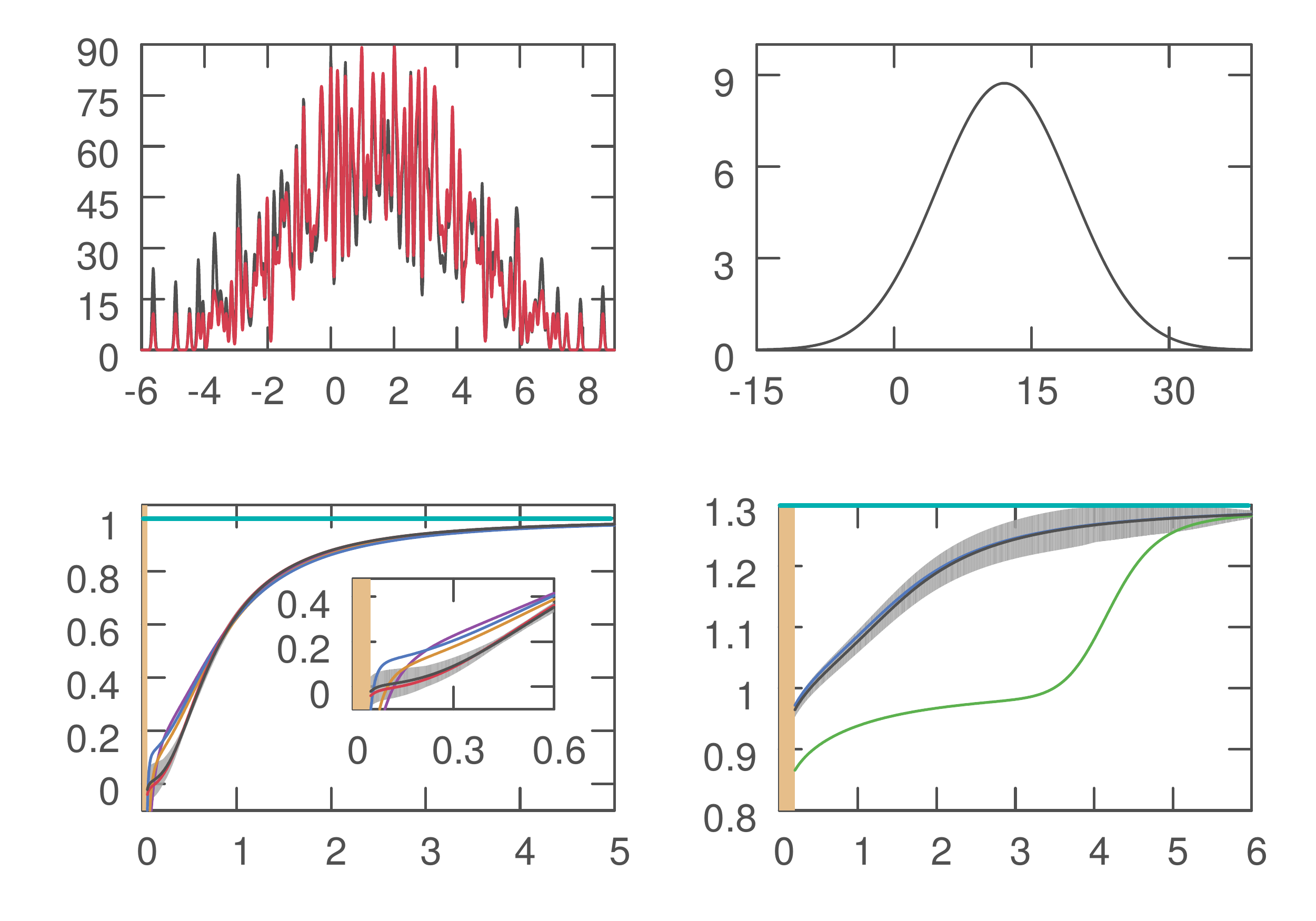}
 \put(-2, 152){$D'_{\eta}$} 
 \put(-2, 65){$\displaystyle \frac{S}{L}$} 
 \put(120, 152){$D'_{\eta}$} 
 \put(120, 65){$\displaystyle \frac{S}{L}$}
 \put(100, 87){$E$} 
 \put(216, 87){$E$} 
 \put(223, 0){$T$} 
 \put(104, 0){$T$} 
 \put(143, 166){\scriptsize $\times 10^{26}$}
 \put(92, 166){\footnotesize $L=6$}
 \put(206, 166){\footnotesize $L=50$}
 \end{overpic}
\end{center}
\caption{ \label{fig:hubbdos} (color online) Hubbard model for $U=J$. 
Upper left: broadened DOS $D_\eta' $ for $L=6$, exact (red curve) and calculated via HS-TN for $m=64$ and $N_R=50$ (black curve). 
Upper right: $D_\eta' $ for $L=50$, via HS-TN with $m=30$, $N_R=20$.
Lower left: Entropy per site $S/L$ for $L=6$, $N_R=50$ and $m=10$ (purple), $m=30$ (orange), $m=50$ (blue), $m =64$ (black), exact (red).  
Lower right: Entropy per site for $L=50$ and $N_R=50$ and $m=10$ (green), $m=25$ (blue), $m=30$ (black).  
Cyan lines represent the analytical limit for high temperature. 
The Insets show a magnification of the bigger panels in the low-temperature regime.
The gray areas represent the estimated error from the ensemble average over $N_R$, while the orange areas 
highlight our temperature threshold $T_{\text{lim}} \approx \eta^{-1/2}$.
} 
\end{figure}

To demonstrate that the HS-TN method is not limited to the study of free-particle models, we apply it to the study of the spin-$\frac{1}{2}$ Fermi-Hubbard model, $H = U \sum_j^{L} n_{\uparrow,j} n_{\downarrow,j} - J \sum_{j}^{L-1} \sum_{\sigma} c^{\dagger}_{\sigma,j} c_{\sigma,j+1} + \text{h.c.}$, restricted to the symmetry sector of $N_{\uparrow} = L/2$ spin-up particles and $N_{\downarrow} = L/2$ spin-down particles (addressing explicitly the U(1) $\times$ U(1) subgroup of the full symmetry group for the model).
The $D'_{\eta}(E) := D_{\eta,N_{\uparrow} = N_{\downarrow} = L/2}(E)$ are plotted in Fig.~\ref{fig:hubbdos} 
for different system sizes: $L=6$ which can be exactly diagonalized, and $L=50$ where convergence is observed. 
In the lower panels of Fig.~\ref{fig:hubbdos}, we report the entropy per site $S/L = -L^{-1} \partial \free / \partial \temp$, 
again confirming convergence to the exact solution where available.
As a side note, we remark that when we adopt a broadening $\eta$ able to resolve the energy gap (such as in Fig.~\ref{fig:hubbdos} for $L=6$),
we can in turn access temperatures sufficiently low to display the insulating behaviour (signalled by the plateau in the bottom-left panel).

\textit{Spectral quantities} $-$
The HS-TN method is not limited to the calculation of the DOS,
but it can be applied to compute the spectral properties related to arbitrary observables $\Theta$.
It is possible to calculate the expectation value $\bar\Theta_{\eta}(E) = \trace[\Theta \hat{\ense}_\eta(E)]$ 
over the broadened microcanonical ensemble $\hat{\ense}_\eta(E) := \bDOS^{-1}(E) \sqrt{\frac{\eta}{\pi}} e^{-\eta(E-H)^2}$.
This allows to calculate the thermodynamical properties
$\bar\Theta(T) = e^{\free_{\eta} / \temp} \int e^{-E/T} \bar{\Theta}_{\eta}(E) \bDOS(E) dE$ for temperatures $\temp \gg \eta^{-1/2}$.
Analogously to Eq.~\eqref{eq:bdosintegral}, the spectral quantity can be recast into the integral
\begin{equation} \label{eq:spectralfun}
 \bar{\Theta}_{\eta}(E) = \frac{1}{\pi \bDOS(E)} \int_{0}^{+\infty} e^{-\frac{t^2}{4\eta}} \,\mathfrak{Re}\!\left( e^{itE} \,\trace\!\left[ \Theta e^{-itH}\right] \right) dt,
\end{equation}
via the HS equation.
In this picture, numerically calculating $\trace[ \Theta e^{-itH} ]$ using TEBD requires little additional effort
than calculating simply $\trace[ e^{-itH} ]$, 
provided that $\Theta$ can be efficiently expressed as a Matrix Product Operator \cite{MPOreps}.
In practice, this requires the computation of the delayed matrix elements
$\langle \psi_R (0) | \Theta |\psi_R (t) \rangle$ for pathway {\it(1)}, since
$\trace\!\left[  \Theta e^{itH}\right] = \frac{d^{L}}{N_R} \sum_{R}^{N_R}\langle \psi_R (0) | \Theta |\psi_R (t) \rangle$,
or alternatively of
$\langle \Phi^+|^{\otimes L} (\Theta e^{-itH})_{\text{odd}} \otimes \Id_{\text{even}} | \Phi^{+} \rangle^{\otimes L} = \trace[\Theta e^{-itH}]$
for pathway \textit{(2)}.

\begin{figure}
 \begin{center}
 \begin{overpic}[width = \columnwidth,  unit=1pt]{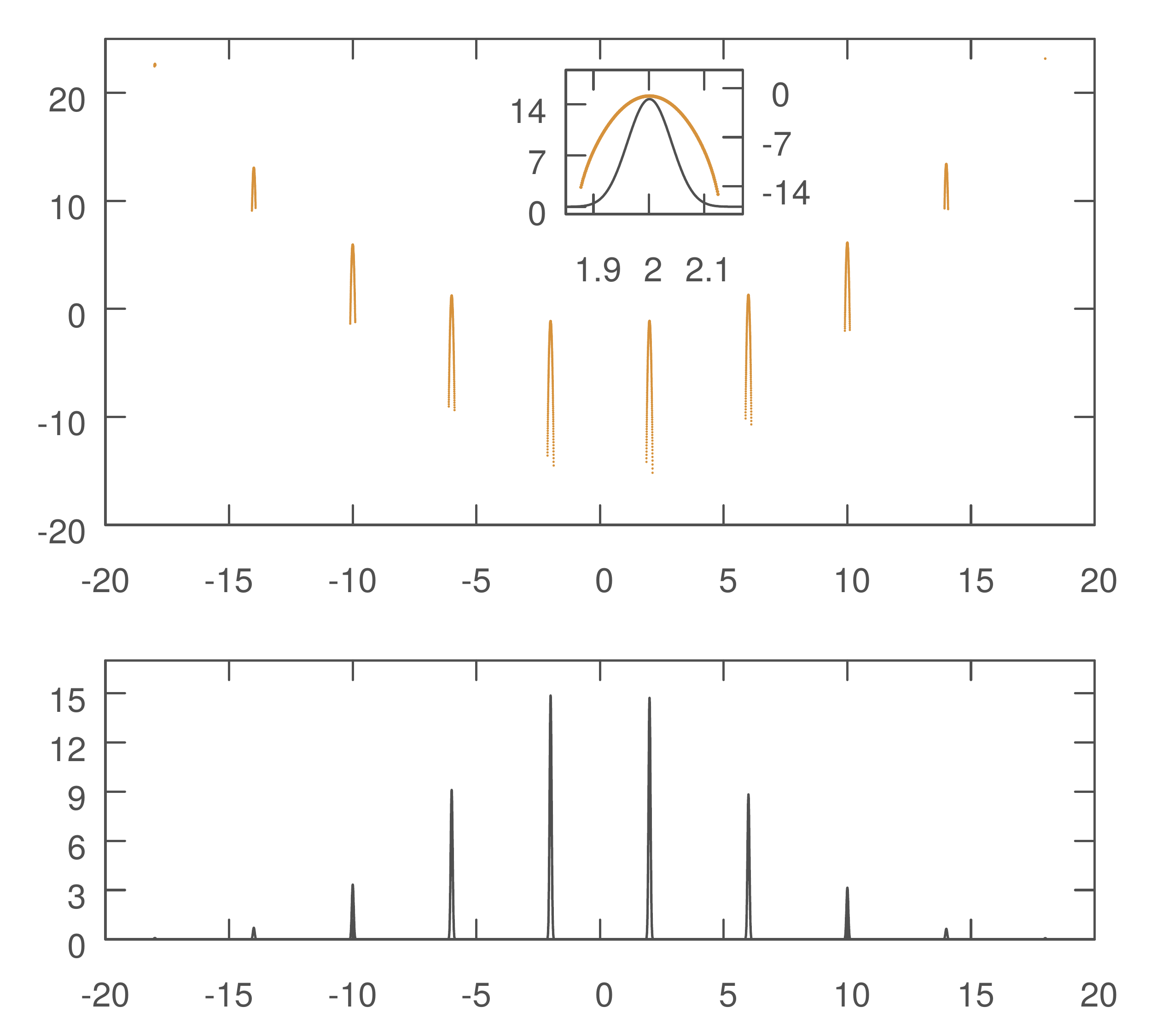}
 \put(217, -2){$E$} 
 \put(-3, 80){$D_{\eta,+}$} 
 \put(217, 87){$E$} 
 \put(-5,217){$\langle \sigma^z_{j} \sigma^z_{j+1} \rangle_{\eta,+}$}
 \put(21, 82){\scriptsize $\times 10^{8}$}
 \put(235, 19){\scriptsize $\times 10^{2}$}
 \put(235, 108){\scriptsize $\times 10^{2}$}
 \put(172, 196){\scriptsize $\langle \sigma^z_{j} \sigma^z_{j+1} \rangle_{\eta,+}$}
 \put(90, 194){\scriptsize $D_{\eta,+}$}
 \put(106, 206){\tiny $\times 10^{8}$}
 \put(157, 165){\tiny $\times 10^{2}$}
 \end{overpic}
\end{center}
\caption{ \label{fig:spectralq} (color online)
Spectral function of the magnetization n-n correlations in the paramagnetic direction $\sigma^{z}_{j} \sigma^{z}_{j+1}$ (orange points), averaged over pairs $j$, plotted
as a function of the energy $E$, where we considered the even sector $+$ of the Ising model at $h = 100$ on $L = 30$ sites.
The inset shows a magnification of both data sets (using the same color code) focusing around $E = 200$, using two different scales for the $y$-axis.
The bottom plot shows the corresponding DOS $D_{\eta,+}$ (black curve), where
we chose a broadening ($\eta = 9000$) far smaller than the width of the visible peaks.
Data have been obtained with the HS-TN method, using bondlink $m = 100$ and $N_R = 20$ initial random samplings.
} 
\end{figure}
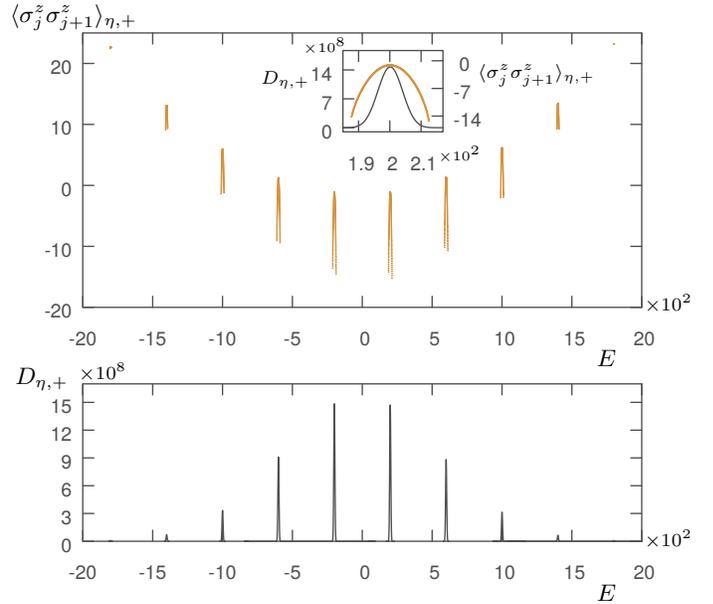

Similarly, it is possible to reconstruct spectral probability distributions
$p_\eta(E) = D_{\eta}(E) \left\langle \Psi | \hat{\rho}_{\eta}(E) | \Psi \right\rangle$ for a given MPS $|\Psi\rangle$.
Figure \ref{fig:spectralq} shows an example of such treatment, performed on the even sector of the Ising model,
on which we spectrally probe the nearest-neighbour correlations of the paramagnetization $\langle \sigma^{z}_j \sigma^{z}_{j+1} \rangle$.

%\textit{Conclusions} $-$
\section{Conclusions and Outlook}\label{sec:conc}

We introduced the HS-TN method to reconstruct the complete broadened microcanonical ensemble of one-dimensional quantum 
many-body models. It relies on performing real short-time evolution of Matrix Product States, which can be performed via standard TEBD algorithm, 
and then manipulates the information of the dynamics into spectral quantities by the Hubbard-Stratonovich transformation.
The method we discussed has the advantage of accessing system sizes far beyond the exact diagonalization, while at the same
time capable of simulating systems at finite densities or fermionic systems without incurring into any sign problem. Finally,
we remark that techniques for performing time-evolution of MPS (such as TEBD or TDVP) are well established and largely
available as free-released software, ultimately making our method a ready-to-use toolkit.
We benchmarked the method by explicitly calculating the broadened density of states and some 
corresponding thermodynamical quantities for free and interacting fermion models. 

We conjecture that the good accuracy of the HS-TN method even with relatively small bond dimensions $m$ is due to mixing effects. 
Indeed, on the one hand, high-energy eigenstates might be highly-entangled. On the other hand, 
it is often possible to express mixtures (in this case Gaussian ensembles) of states (at fixed resolution $\eta$, in numbers exponentially increasing with $L$) with a relatively small amount of quantum correlations, and thus to represent such mixtures efficiently with a tensor network. 
In fact, this effect has been already observed and discussed in other scenarios \cite{MPDO,LPTN,Gemma,METTS,METTS2}.
More quantitative observations regarding such behaviour are reported in Appendix~\ref{app:observe}.

We showed explicit examples of the HS-TN technique applied to one-dimensional models in open boundary conditions. However, we stress that this method extends to all those models whose out-of-equilibrium, short timescale dynamics can be numerically simulated efficiently and precisely, thus effectively extending the scope of our treatment beyond 1D.

This manuscript opens wide new perspectives in the framework of tensor networks methods, as we have shown that they are
suitable tools to study complex quantum systems well beyond the low-temperatures limit. The HS-TN technique 
can be applied to a plethora of cases, e.g. the evaluation of thermodynamical and transport properties 
of quantum systems (spin chains, fermionic and bosonic models, nanotubes, small molecules, etc.) whose 
detailed knowledge has been so far precluded.

\textit{Acknowledgements} $-$
The authors acknowledge support from the EU via RYSQ, UQUAM and QUIC projects, the DFG via SFB/TRR21 project, the Baden-W\"urttemberg Stiftung via Eliteprogramm for Postdocs and the Carl Zeiss Foundation.
SM gratefully acknowledges the support of the DFG via a Heisenberg fellowship.
RF kindly acknowledges support from the National Research Foundation of Singapore (CRP - QSYNC) and the Oxford Martin School.

%%%%%%%
% After running bibtex for the last time, perform the .bbl substitution
%%%%%%%
%\bibliography{HSbiblio}

\appendix
\section{Parameters and computational complexity} \label{app:finetune}

We now discuss in detail fine-tuning issues of the simulation parameters and the related computational complexity.
%Moreover, we show numerical convergence of pathway {\it (1)} (random initial state sampling) in the number of random samplings $N_R$. Finally, we investigate a scaling of the algorithm performance with the system size.
Hereafter, we consider a MPS ansatz of bond dimension $m$ for the real-time dynamics, on a 1D system of size $L$ with open boundary conditions.
The parameter $\eta$ introduces a spectral broadening of magnitude $\delta E \sim \eta^{-1/2}$
We numerically evaluate the HS-integral given in Eq.~(2) by introducing a discrete time-step
$\mathrm{dt}$ and finite evolution-time bound $\bar{t}$ reducing the integration interval $\int_{0}^{\infty}$ to $\int_{0}^{\bar{t}}$.
We tune the integration parameters according to
\begin{equation}
\begin{aligned}
\mathrm{dt} & \simeq  {\Delta E }^{-1}\\ 
\bar{t}     & \simeq  \sqrt{4\eta} \, \mathrm{erfc}^{-1}\left(\chi\right), 
\end{aligned}
\end{equation}
where $\Delta E = E_\mathrm{max} - E_\mathrm{min}$ is the estimated width of the complete spectrum, and
$\mathrm{erfc}(z) = \frac{2}{\sqrt{\pi}} \int_{z}^{\infty} e^{-t^2} dt$ is the complementary error function. The additional fine-tuning parameter $\chi \ll 1$ controls the truncation-error of the Gaussian modulation (we typically set  $\chi$ of the order of $10^{-10}$). 

As the models we typically consider are short-range, we can safely assume a linear bound with $L$ to the spectral width
$\Delta E \propto L$.
Under this condition, we observed that we can employ dt as time-step for the TEBD algorithm in order to
obtain a target precision independent of $L$.
Consequently, the computational complexity of the algorithm equals the cost of performing ${\cal N}$ time-steps of TEBD, where
${\cal N} = \bar{t} / \mathrm{dt} \propto \Delta E / \delta E \propto \sqrt{\eta} L$. As a single TEBD time-step carries
a cost of the order $\mathcal{O} \left( L m^3 \right)$
[21$-$23] %\cite{VidalTEBD,daley04,TDMRG}
(including measurements), the overall cost of the HS-TN algorithm with MPS is

\begin{equation}
\label{eq:complexity}
\text{Complexity of HS-TN} = \mathcal{O} \left( \eta^{\frac{1}{2}} \, L^2 m^3 \right).
\end{equation}
The computation of the $\bDOS(E)$ using $\mathcal{O}(\mathcal{N})$ points $E\in \left[E_\mathrm{min},E_\mathrm{max}\right]$ is simply a (fast) Fourier-transformation,
of marginal computational cost.

Two noteworthy remarks follow: First, the bond dimension $m$ is expected to increase with $L$, however our test-cases shows that a quite small $m$ over various system-sizes is sufficient to obtain accurate results.
Second, in pathway {\it (1)}, the random sampling introduces an additional factor $N_R$ to the overall cost.
However, we observed that often we do not need to increase $N_R$ for larger $L$ to maintain the fixed precision
(on the contrary, the precision can even improve with $L$ at fixed $N_R$).

\begin{figure}
 \begin{center}
 \begin{overpic}[width = \columnwidth, trim={0pt 10pt 0pt 0pt},clip, unit=1pt]{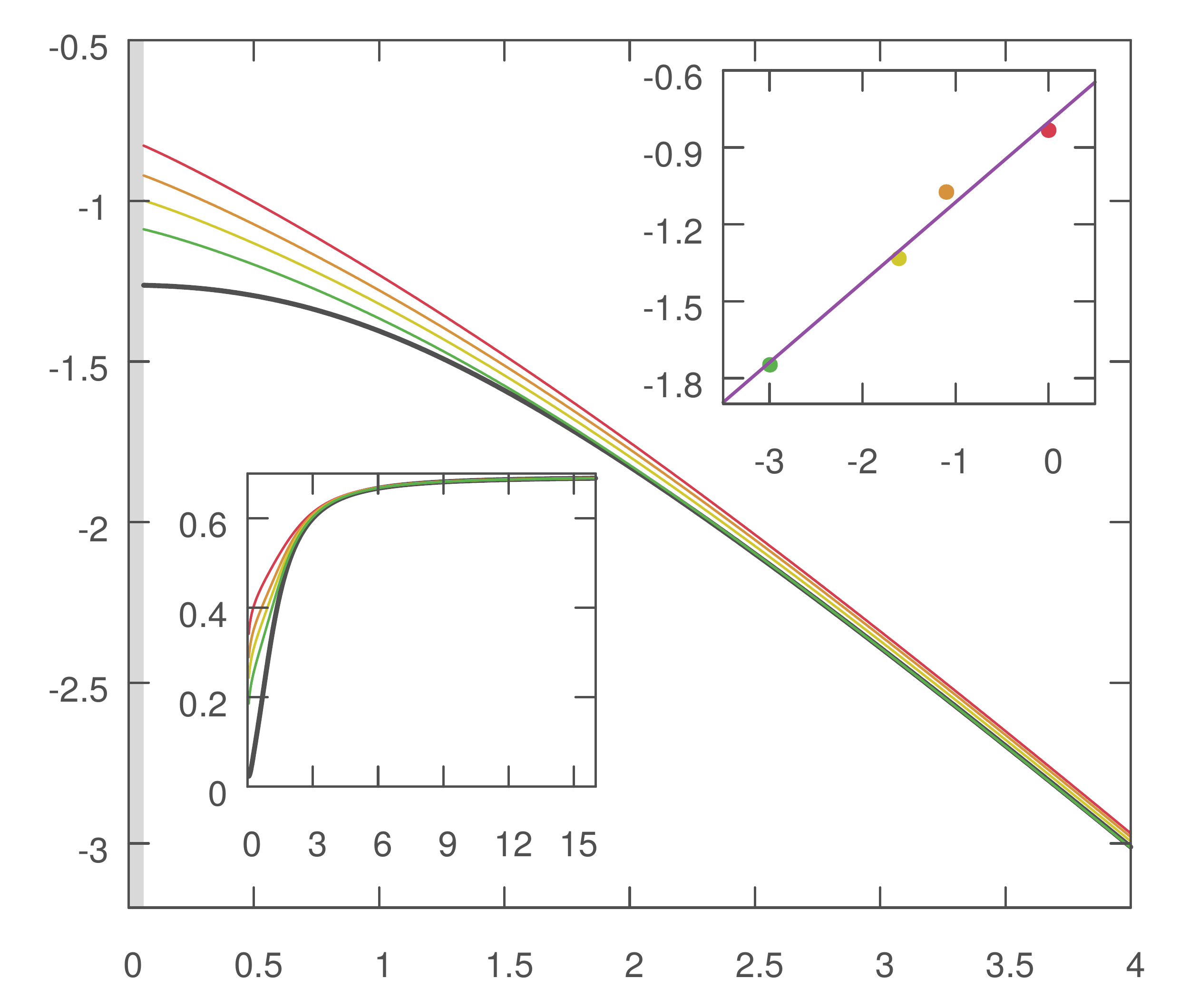}
 \put(0, 165){$\displaystyle\frac{\free}{L}$}
 \put(193, -4){$T$}
 \put(39, 112){$S/L$}
 \put(129, 30){$T$} 
 \put(110, 165){$\log \epsilon'$}
 \put(182, 95){$\log (1/N_R)$}
 \end{overpic}
%
 %\includegraphics[width = \columnwidth]{Figure1.jpg}
 %
 %Sample figure
\end{center}
\caption{ \label{fig:A2} (color online)
Convergence of the HS-TN method with the number of repetitions $N_R$, for the Ising model.
Main panel: Free energy per site $\free/L$ as a function of the temperature $T$, for $N_R=1$ (red), 3 (orange), 5 (yellow) and 20 (green). The black curve shows the exact (broadened) result; the shaded grey area on the left indicates the temperatures regime where our results are not stable. Bottom inset: Entropy per site $S/L$, using the same color code.
Top inset: maximal error $\epsilon'$ between $\free/L$ calculated with HS-TN method and the exact solution, plotted as a function of the number $N_R$ of random initial MPS, in double logarithmic scale. The purple line shows a power-law fit, corresponding to a power of $\sim 0.3$.
For these plots we used $L=30$, $m=180$, $\eta = 10 L^2$, $h = 1$, and the even symmetry sector was considered.
}
\end{figure}
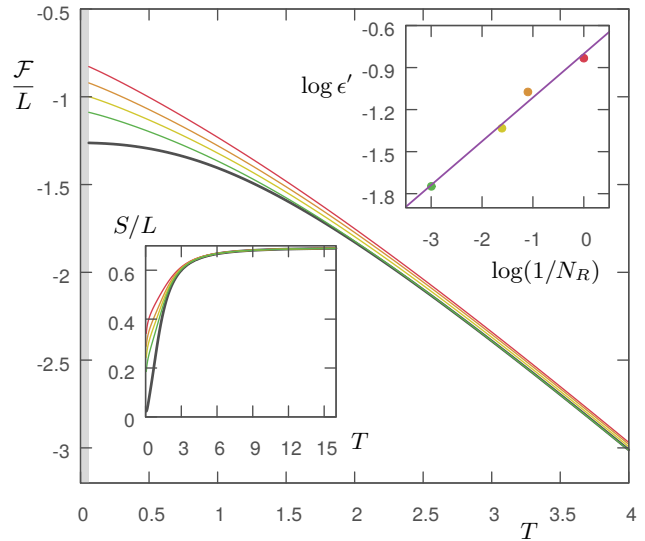

\section{Convergence with $N_R$ and generation of symmetric random states} \label{app:repeat}

\begin{figure}
 \begin{center}
 \begin{overpic}[width = \columnwidth, unit=1pt]{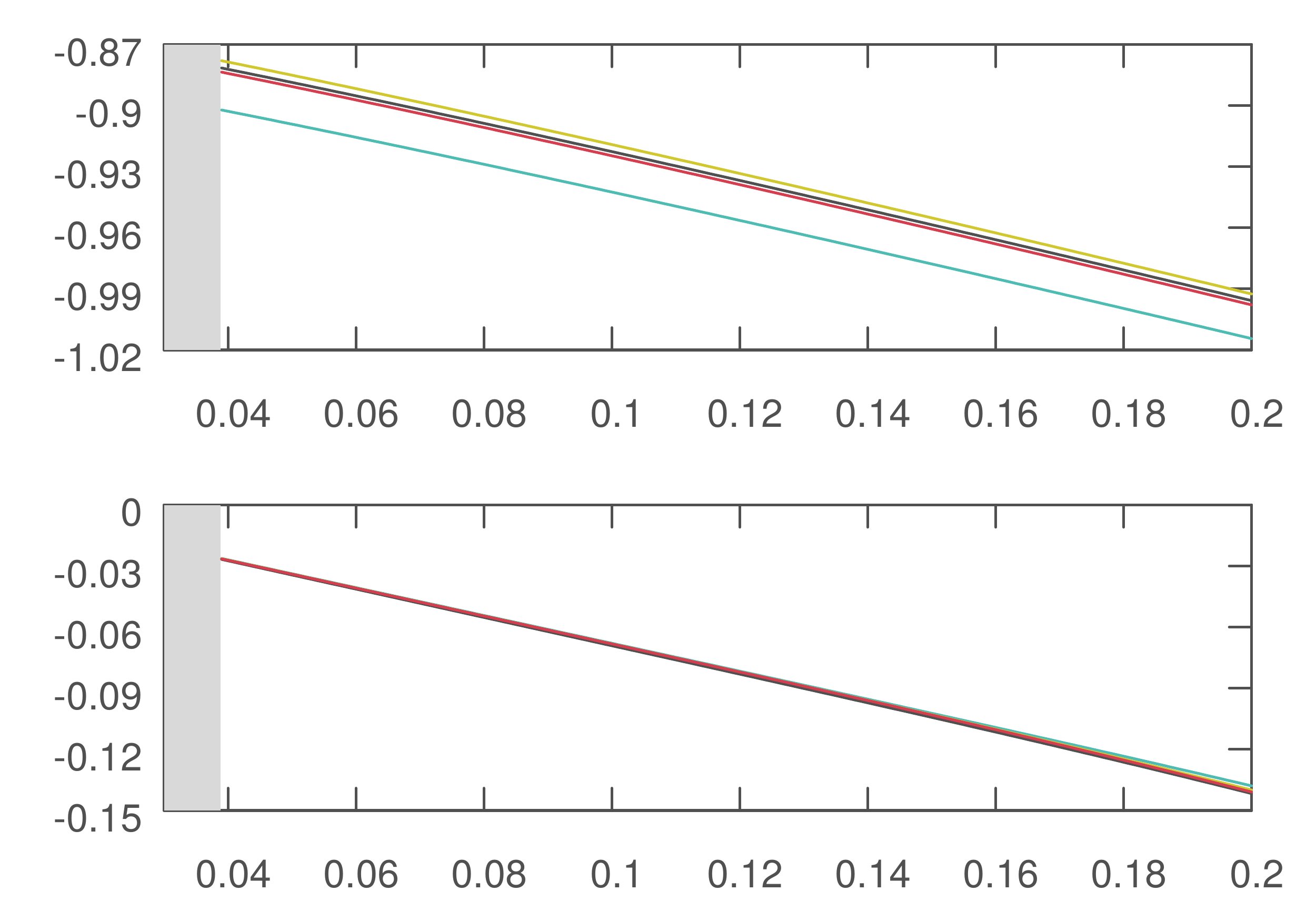}
 \put(-5, 90){$\displaystyle\frac{\free}{L}$} 
 \put(214, -4){$T$} 
 \end{overpic}
\end{center}
\caption{ \label{fig:A3} (color online)
Convergence of the HS-TN method with the number of repetitions $N_R$, for the Hubbard model.
Top panel: Weakly insulating regime  $J/U=10^3$. Bottom panel: Strongly insulating regime $J/U = 0$.
Both panels show the free energy per site $\free/L$ plotted as a function of the temperature $T$, for $N_R=1$ (cyan), 5 (yellow), 15 (black) and 20 (red).
Here $L=30$ and $m=180$, and $\eta = 10 L^2$ were used.
}
\end{figure}

Here we report the convergence of the ensemble reconstruction using pathway {\it (1)} (random initial state sampling), as a function of the number $N_R$ of initial random states employed.
To properly test convergence with  $N_R$, we consider results at convergence in the other refinement parameters (e.g.~$m$).
Figures \ref{fig:A2} and \ref{fig:A3} show convergence of the free energy calculated via HS-TN method, as a function of $N_R$, for the Ising model (even symmetry sector) and the Hubbard model (half-half filling symmetry sector), respectively.
In the Ising model, we observe once more a power-law convergence to the exact result of the error $\epsilon'$ as a function of $1/N_R$.

In all cases, the initial random MPS with bond dimension $m$, and quantum number $q$ of a global Abelian symmetry, were sampled as follows: We randomly draw quantum numbers $q$, for each bond-index $1,\dots,m$ and each bond, from the maximal set of all nontrivial contributions. This is equivalent to picking a random $m$-dimensional sub-representation of the largest possible symmetry representation for each bond link.
Then,  as an MPS in a fixed global sector $q$ has all block-diagonal tensors, we sample the elements of the nontrivial blocks from a normal distribution.
Such sampled states were then normalized and gauged properly for the TEBD evolution.

\section{Resolution and scaling with $L$} \label{app:observe}

\begin{figure}
 \begin{center}
 \begin{overpic}[width = \columnwidth, trim={0pt 0pt 0pt -18pt},clip, unit=1pt]{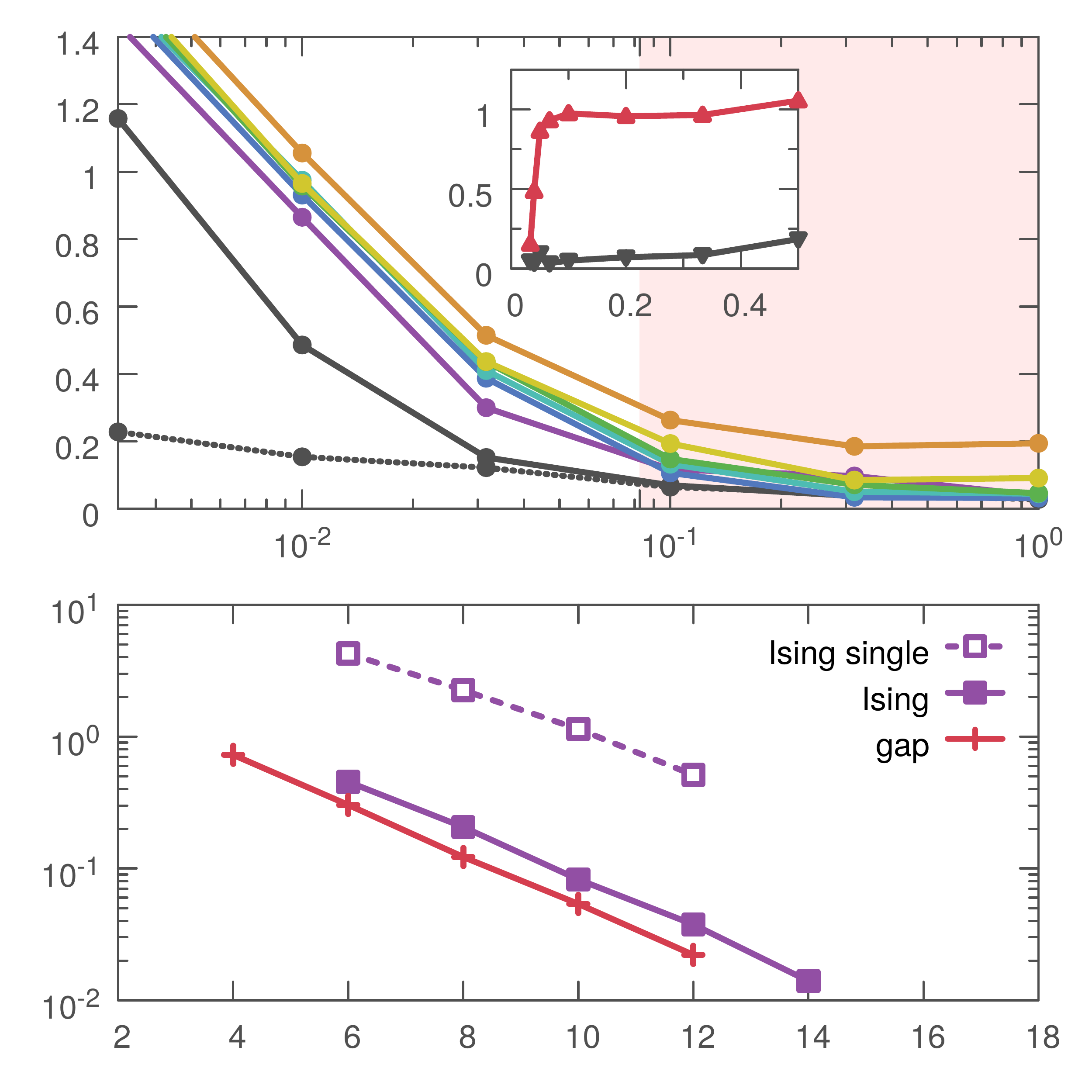}
 % main
 \put(0, 228){$\epsilon_0$}
 \put(214, 116){$\delta E$} 
 \put(28, 241.76){\footnotesize Ising, $L=10$}
 \put(120, 118){$\overline{\delta E}$}
 \put(133, 125){\vector(2,1){10}}
 % main inset
 \put(94, 223){$\epsilon_0$}
 \put(178, 172){$1/m$}
 \put(184, 191){$\delta E > \overline{\delta E}$}
 \put(184, 221.5){$\delta E < \overline{\delta E}$}
 % bottom left
 \put(0, 86){$\overline{\delta E}$}
 \put(220, 0){$L$}
 \end{overpic}
\end{center}
\caption{ \label{fig:resolimit} (color online)
         Accessible spectral resolutions with the HS-TN method. 
	Top: Error $\epsilon_0$ 	between exact broadened DOS and HS-TN DOS, plotted over the selected resolution $\delta E = \eta^{-1/2}$ in the Ising-model at $h=1$, $L=10$.
	The white area, $\delta E > \overline{\delta E}$, shows the regime where $\epsilon_0(m)$ converges rapidly to 0 in $m$ (here we used $m=2$ (orange), $3$ (yellow), $5$ (green), $10$ (cyan), $15$ (blue), $20$ (purple), and $25$ (black), $30$ (black dotted) near saturation, and $N_R = 35$). Conversely, the pink area signals the high resolution regime $\delta E < \overline{\delta E}$ where convergence is hindered.
	Inset: Scaling of $\epsilon_0$ with the (inverse) bond dimension $1/m$ in a low-resolution $\delta E_\mathrm{in}=10^{1/2} > \overline{\delta E}$ (black), and a high-resolution $\delta E_\mathrm{out}=10^{-2} < \overline{\delta E}$ (red) scenario respectively.
	Bottom: Resolution limits $\overline{\delta E}$ estimated via HS-TN respectively with $N_R = 1$ (empty boxes) and $N_R$ at convergence (full boxes), 
	as a function of the system size $L$, compared with the typical energy gap $\sqrt{\left\langle d E_{\nu}^2 \right\rangle}$.
}
\end{figure}
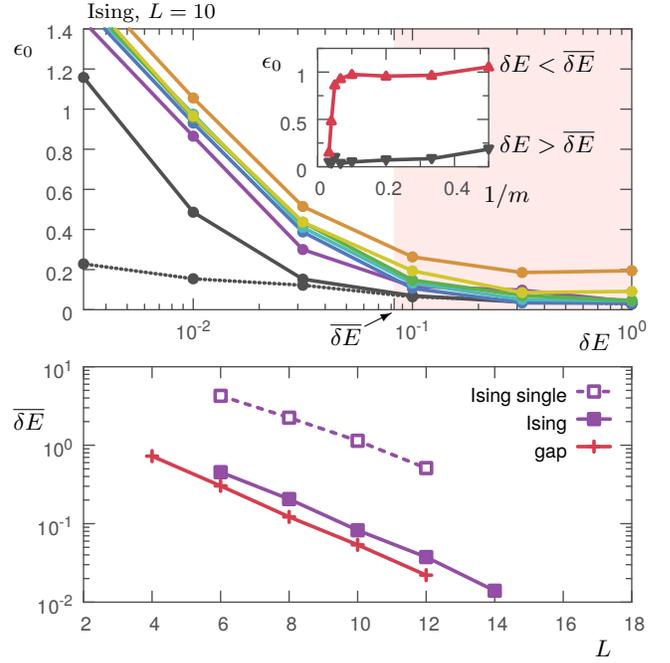

In this section, we now focus on the interplay between the system-sizes $L$ and the broadening-parameter $\eta$.

As a measure of precision, we consider the error $\epsilon_0(m,\eta)$, defined as in Fig.~(1), between the normalized broadened $D_{\eta,+,\text{exact}}$ (exact result) and $D_{\eta,+,m}$ computed with the HS-TN method, over various bond-dimensions $m$ and broadening-parameters $\eta$ in the Ising model. Our results are summarized in Fig.~\ref{fig:resolimit}, and lead to the following observations:
It emerges that for a given system size $L$, there is an inherent broadening threshold $\bar{\eta}(L)$, or equivalently a spectral resolution threshold $\overline{\delta E} = \bar{\eta}^{-1/2}$ accessible via HS-TN. The top panel of Fig.~\ref{fig:resolimit} shows that the error $\epsilon_0(m,\eta)$ converges to zero rapidly in $m$ as long as $\delta E > \overline{\delta E}$ (i.e.~$\eta <  \bar{\eta}(L)$).
Conversely, when a resolution higher than this threshold is selected ($\delta E < \overline{\delta E}$) a non-zero plateau in the error emerges for large $m$, until we reach entanglement saturation of the TN ansatz. This behaviour is highlighted in the inset (red data points).
In turn, in all the cases we considered, such resolution threshold $\overline{\delta E}(L)$ improves with the system size, i.e.~it decreases exponentially with $L$ (bottom panel).

We interpret this behaviour with an argument based on the energy spacings of the Hamiltonian $H$, i.e.~the differences between subsequent eigenvalues of $H$: It appears, in fact,
that the spectral resolution threshold is well approximated by the typical energy spacing $\overline{\delta E} \approx \Delta E / d^{L}$ 
(or more precisely $\overline{\delta E} \approx \sqrt{\left\langle d E_{\nu}^2 \right\rangle}$, compare red and purple data sets in the bottom plot).
If the spectral broadening we introduce is larger than the typical energy-spacing, then
we do not need to reconstruct single eigenstates of $H$ (expensive to represent with tensor networks when they exhibit a volume-law
of entanglement) but mixtures of them, which instead may be efficiently approximated by TN.
According to this argument,
the HS-TN method will converge rapidly in $m$ if we set $\delta E > \overline{\delta E}$, as observed above.

Finally, since the total spectral width is usually extensive $\Delta E \propto L$, the average energy gap decays roughly exponentially with $L$
($\overline{\delta E} \propto e^{-L}$). From this follows that, as the thermodynamical limit is approached $L \to \infty$, the resolution threshold vanishes $\overline{\delta E} \to 0$, 
and we expect the HS-TN method to converge regardless of the chosen broadening.

\newpage

\end{document}